\documentclass{aa}
\usepackage{times}
\usepackage{latexsym}
\usepackage{epsfig}
\def\kms{km~s$^{-1}$}
\def\0{\hspace*{0.5em}}
\def\vinf{$v_{\infty}$}

\begin{document}

\title{On the absence of wind bow-shocks around OB-runaway stars:
probing the physical conditions of the interstellar medium}

\author{Fredrik Huthoff\inst{1} \and Lex Kaper\inst{1}} 

\offprints{L. Kaper;  e-mail: lexk@science.uva.nl}

\institute{Astronomical Institute ``Anton Pannekoek'' and Center for
           High-Energy Astrophysics, 
           University of Amsterdam, \\ Kruislaan 403, 1098 SJ Amsterdam, 
           The Netherlands}

\date{Received; Accepted}

\authorrunning{F. Huthoff \& L. Kaper}
\titlerunning{Wind bow-shocks as probes of the interstellar medium}

\abstract{High-resolution IRAS maps are used to search for the
presence of stellar-wind bow-shocks around high-mass X-ray binaries
(HMXBs). Their high space velocities, recently confirmed with {\it
Hipparcos} observations, combined with their strong stellar winds
should result in the formation of wind bow-shocks. Except for the
already known bow-shock around Vela~X-1 (Kaper et al. \cite{kaper97}),
we do not find convincing evidence for a bow-shock around any of the
other HMXBs. Also in the case of (supposedly single) OB-runaway stars,
only a minority appears to be associated with a bow-shock (Van Buren
et al. \cite{vanburen95}).\\ We investigate why wind bow-shocks are
not detected for the majority of these OB-runaway systems: is this due
to the IRAS sensitivity, the system's space velocity, the stellar-wind
properties, or the height above the galactic plane? It turns out that
none of these suggested causes can explain the low detection rate
($\sim$40~\%). We propose that the conditions of the interstellar
medium mainly determine whether a wind bow-shock is formed or not. In
hot, tenuous media (like inside galactic superbubbles) the sound speed
is high ($\sim$100~\kms), such that many runaways move at subsonic
velocity through a low-density medium, thus preventing the formation
of an observable bow-shock. Superbubbles are expected (and observed)
around OB associations, where the OB-runaway stars were once
born. Turning the argument around, we use the absence (or presence) of
wind bow-shocks around OB runaways to probe the physical conditions of
the interstellar medium in the solar neighbourhood.
}

\maketitle

\section{Introduction}

OB-runaway stars are massive, OB-type stars with high space
velocities. This small ($\sim$20~\% of the O stars, less than 5~\% of
the early B stars), but significant excess of high-velocity objects
suggests a systematical production of fast movers (Feast \&
Shuttleworth \cite{feast}, Stone \cite{stone}). Among the OB runaways,
velocities well above 100~\kms\ have been observed (Gies \& Bolton
\cite{gies86}, Conlon et al. \cite{conlon}, Blaauw \cite{blaauw93}),
which is roughly ten times the average space velocity of ``normal'' OB
stars in the Milky Way. Due to their high velocities, OB-runaways may
have travelled great distances since their formation. The criteria
used to classify OB stars as ``runaways'' are (Blaauw
\cite{blaauw61}): (i) A high space velocity; often a threshold space
velocity of 30~\kms\ is used, roughly corresponding to three times the
typical space velocity of OB-type stars. (ii) The reconstructed path
of the runaway should start in a ``parent'' OB association.  Since the
definition by Blaauw, the term {\it OB runaway} has often been used to
include all of the high-velocity OB stars, irrespective of the
identification of a parent OB-association. A large distance above the
galactic plane can also be used as an indication for the
runaway-nature of a massive star (e.g. Van Oijen \cite{vanoijen}).

The two most popular scenarios for the production of runaways seem to
operate at roughly the same rate (Hoogerwerf et al. \cite{hoogerwerf}) and
are both capable of producing runaways with velocities up to 200~\kms\
(Portegies-Zwart \cite{portegies}, Leonard \cite{leonard}): (i) The
{\it binary supernova scenario} (Blaauw \cite{blaauw61}, Zwicky
\cite{zwicky}), where a massive star in a binary receives a high
space velocity after the supernova explosion of its (initially) more
massive companion; (ii) The {\it cluster ejection scenario} (Poveda et
al. \cite{poveda}), where dynamical interactions in a compact cluster
result in the ejection of one or more of its members. Especially
encounters between binary systems are effective in producing
high-velocity objects (e.g. Mikkola \cite{mikkola}).

The current version of the binary supernova scenario includes a phase
of mass transfer (Van den Heuvel \& Heise \cite{VdhH72}): when the
initially most massive star fills its Roche lobe (e.g. when becoming a
supergiant) mass is transfered to its companion which eventually will
become the most massive star in the system. As a consequence, the
system has a high probability to remain bound after the supernova
explosion of the initially most massive star (Boersma
\cite{boersma61}). If it remains bound, the binary, now consisting of
a massive, rejuvenated main-sequence star and a compact remnant (a
neutron star or a black hole), will travel through space with a high
velocity. As soon as the massive star becomes a supergiant, material
from its dense stellar wind (or through Roche-lobe overflow) is
intercepted by the gravitational field of the compact star and
accretes, powering a strong X-ray source: a high-mass X-ray binary
(HMXB). This scenario thus predicts that all HMXBs are runaway
objects. Recently, this prediction has been confirmed with {\it
Hipparcos} observations (Chevalier \& Ilovaisky \cite{chevalier},
Kaper et al. \cite{kaper99}). This also suggests that several
supposedly single OB runaways might have a, so far undetected, compact
companion. For a review on HMXBs and OB-runaway stars, see Kaper
(\cite{kaper01}).

\begin{table*}
\centering
\caption{Prediction of the angular separation between wind bow-shock
and HMXB. The HMXB peculiar velocity (i.e. corrected for differential
galactic rotation and peculiar solar motion) in the radial and
tangential direction is listed in column 6 and 7, respectively; the
resulting space velocity $v_{\rm space}$ is given in column 8. The
first five columns list the identification of the X-ray source, the
name and spectral type of the OB companion (if known), the distance
$r$, and the height $z$ above the galactic plane. The predicted (and
observed, final column) angular distance between OB star and wind
bow-shock are given in the last columns. For the HMXBs that have not
been observed with {\it Hipparcos} $(v_{\rm tan})_{pec}=$ 50~\kms\ has
been adopted. If a question mark is added, infrared emission is
detected, but no clear arclike structure could be resolved. The
prediction of the stand-off distance is based on stellar-wind
parameters taken from Puls et al. (\cite{puls}) and Kaper et
al. (\cite{kaper98}); for Be stars it was assumed that $\dot{\rm M} =
10^{-8} {\rm M}_{\odot}~{\rm yr}^{-1}$ and $v_{\infty}$ = 250~\kms.}
\renewcommand{\arraystretch}{1.4} \setlength\tabcolsep{5pt}
\begin{tabular}{lllcrrcc|cc|cc}
\hline\noalign{\smallskip}
Object & & Spectral Type& $r$ & $z$ & $(v_{\rm rad})_{pec}$ & 
$(v_{\rm tan})_{pec}$ & $v_{\rm space}$ & $d_{\rm shock}^{\rm pred}$ & 
$\delta_{\rm shock}^{\rm pred}$ & bow-shock? & $\delta_{\rm shock}$ \\
& & & (kpc) & (pc) & (\kms) &(\kms) &(\kms) &(pc) & $(')$ & & $(')$   \\
\noalign{\smallskip}
\hline
\noalign{\smallskip}
0114+650 &           & B0.5 Ib   & 3.8  & 171   & -9.6  & 29.4  & 31.0 &  
  2.6  & 2.4 & no & \\   
1223-624 & GX301-2   & B1.5 Ia   & 5.0  & -3.0  & -1.6  & {\em 50} & {\em 50}&  
  2.8  & 1.9 & ? & 1.5(?)  \\
1700-377 & HD 153919 & O6.5 Iaf  & 1.7  & 65    & -42.0 & 57.3  & 71.0 &  
  3.1  & 6.3 & ? & 8.0(?) \\
1907+097 &           & B0.5 Ib   & 4.0  & 37    &       & {\em 50} &{\em 50}&  
  0.6  & 0.5 & ? & 0.5(?)  \\
1538-522 & QV Nor    & B0 Iab    & 5.5  & 206  & -81.9 & {\em 50} &{\em 96} &  
  1.9  & 1.2 & no &  \\
1119-603 & Cen X-3   & O6.5 II-III & 8.0 & 47   & 16.3  & {\em 50} &{\em 53}& 
  1.7  & 0.7 & no & \\
1956+350 & Cyg X-1   & O9.5 Iab  & 2.5  & 135   & -10.9 & 41.4  & 42.8 & 
  5.0  & 6.9 & no & \\ 
0900-403 & Vela X-1  & B0.5 Ib   & 1.8  & 123   & -22.3 & 38.5  & 44.5 & 
  1.4  & 2.7 & yes & 1.0 \\
0236+610 & V615 Cas  & B0e       & 2.0  & 42    & -30.2 & 22.5  & 37.7 &  
  .14     & .24    & no &   \\
0535+262 & V725 Tau  & O9.7e II  & 2.0  & -90   & -41.5 & 57.5  & 70.9 &  
  .07     & .12    & no &   \\
0352+309 & X Per     & O9e III-IV & 0.8 & -235  & -32.0 & 13.4  & 34.7 &  
  .51     & 2.2    & no &   \\ 
\noalign{\smallskip}
\hline
\noalign{\smallskip}
\end{tabular}
\label{tablehmxb}
\end{table*}
 
The high space velocity and powerful stellar wind of the OB-runaway
star will give rise to a strong interaction with the ambient
medium. In the case of supersonic movement, a wind bow-shock is formed
(Baranov et al. \cite{baranov}). The wind bow-shock accumulates gas
and dust from the interstellar medium which is heated by the
ultraviolet radiation field of the OB star. The heated dust
subsequently radiates at infrared wavelengths (Draine \& Lee
\cite{draine}) and the shocked gas becomes visible in strong optical
emission lines like H$\alpha$ and [O~{\sc iii}] 4959,5007~\AA\ (e.g.\
Kaper et al. \cite{kaper97}). Data obtained with the {\it InfraRed
Astronomical Satellite} (IRAS) has been used to search for the
presence of wind bow-shocks around OB runaways by Van Buren \& McCray
(\cite{vanburen88}) and Van Buren et al. (\cite{vanburen95}). They
detected excess infrared (60~$\mu$m) emission for about 30~\% of
the 188 OB stars in their sample. For a significant fraction prominent
arc-like features could be resolved using high-resolution maps
(Noriega-Crespo et al. \cite{noriega}).  When spatially resolved, the
shape of the bow-shock indicates the direction of motion of the
system. The stand-off distance of the shock provides a constraint on
the space velocity, the stellar-wind parameters, and the density of
the ambient medium.

In this study we investigate whether wind bow-shocks are as common for
HMXBs as for (single) OB-runaway stars, by searching for their
infrared emission in high-resolution maps produced from data obtained
with the IRAS satellite. We find that also in the case of HMXBs only a
minority form a wind bow-shock. We investigate whether the detection
of a bow-shock relates to (i) the kinematical properties of the
runaways; (ii) the height above the galactic plane; or (iii) the
spectral type of the OB runaway. The remaining option, namely that the
physical conditions of the ambient medium mainly determine whether a
wind bow-shock is formed, seems to be the most likely one. If true,
this provides the opportunity to use OB runaways as probes of the
physical conditions of the interstellar medium.

\section{A search for bow-shocks around HMXBs}

In our search for the presence of wind bow-shocks around high-mass
X-ray binaries we selected a number of systems for which the
conditions for the formation of a bow-shock seem favourable: a large
space velocity and an OB companion with a strong stellar wind and a
high luminosity. In practice this means that our sample consists
mainly of systems hosting an OB supergiant (cf.\ Kaper
\cite{kaper98}). We also included a few Be/X-ray binaries (with
early-type OBe companion). The HMXB sample is listed in
Table~\ref{tablehmxb}. To determine the size of the area on the sky to
inspect for the presence of a bow-shock, we estimated the angular
distance from the star at which a bow-shock should form. The stand-off
distance ($d_{\rm s}$) is set by the balance in ram pressure between
the stellar wind and the ambient medium (cf.\ Wilkin
\cite{wilkin}). Using typical values for the wind mass-loss rate \.{M}
(in M$_{\odot}$~yr$^{-1}$), wind terminal velocity $v_{\infty}$ and
space velocity $v_{\star}$ (in \kms), and the density (in cm$^{-3}$)
of the ambient medium $n (= \rho / \mu m_{H})$, the stand-off distance
can be written as:
\begin{equation} 
d_{\rm s}= \frac{0.9 \, {\rm pc}}{(v_{\star}/50 \, {\rm km/s})}
\sqrt{\frac{ (\dot{\rm M} / 10^{-6} \, {\rm M}_{\odot}/{\rm yr})
\times (v_{\infty} / 10^{3} \, {\rm km/s})}{n / {\rm cm}^{-3}}}
\end{equation} 
We calculated $d_{\rm s}$ for the systems in our sample
(Table~\ref{tablehmxb}). Stellar-wind parameters were taken from Puls
et al. (\cite{puls}) and Kaper et al. (\cite{kaper98}). The space
velocity is obtained from the observed {\it Hipparcos} proper
motion and radial velocity, corrected for differential galactic
rotation and peculiar solar motion (cf.\ Moffat et
al. \cite{moffat}). The interstellar medium density was computed with
the mean density model of Dickey \& Lockman (\cite{dickey}). The
typical area on the sky that we inspected for the presence of a
bow-shock is $0.5^{\circ} \times 0.5^{\circ}$.

\subsection{HIRAS maps of the HMXB environment}

We used the program {\tt HIRAS} (Bontekoe et al. \cite{bontekoe}) to
produce high-resolution IRAS maps of the circumstellar environment of
the HMXB sample, to inspect the area for the presence of wind
bow-shocks. The IRAS satellite scanned almost the whole infrared sky in
four wavelength bands (12, 25, 60, and 100~$\mu$m) in strips of about
$6'$ wide. The {\tt HIRAS} software takes advantage of the fact that a
given part of the sky has been observed several times, resulting in strips
that partly overlap. Combining these observations then yields a
significant enhancement in spatial resolution. For details of the {\tt
HIRAS}-algorithm and the used method we refer to Bontekoe et al.
(\cite{bontekoe}). The resolution of the images depends on the local
sampling pattern and, therefore, varies with position on the sky; in
our {\tt HIRAS} images it typically is in the range from $0.4'$ to
$1.7'$.

\begin{figure}[t]
\centerline{\psfig{figure=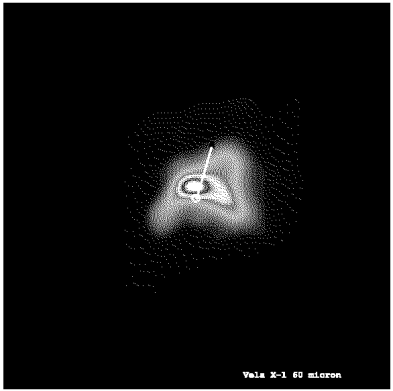,height=7.5cm}}
\caption[]{{\tt HIRAS} image of the circumstellar environment of
Vela~X-1 at 60~$\mu$m. The arc-like structure of a wind bow-shock is
clearly resolved. The arrow indicates the direction of the system's
space velocity. East is to the left and North is up; the image size is
$0.5^{\circ} \times 0.5^{\circ}$. Infrared emission of the bow-shock is
also detected in the other three IRAS bands.}
\label{figvela}
\end{figure}

\begin{figure}[t]
\centerline{\psfig{figure=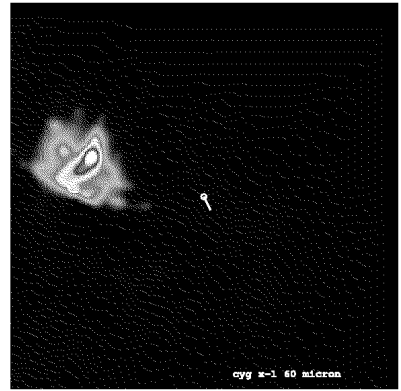,height=7.5cm}}
\caption[]{{\tt HIRAS} image of the environment of Cygnus X-1 at
60~$\mu$m. The feature located roughly $20'$ east of Cyg X-1 is part
of a larger filament that outlines the Cygnus~X region (Cash et
al. 1980): a large X-ray emitting region possibly heated by numerous
supernovae and winds from massive stars. East is to the left and North
is up, the image size is $1.0^{\circ} \times 1.0^{\circ}$. }
\label{figcyg}
\end{figure}

\begin{figure}[t]
\centerline{\psfig{figure=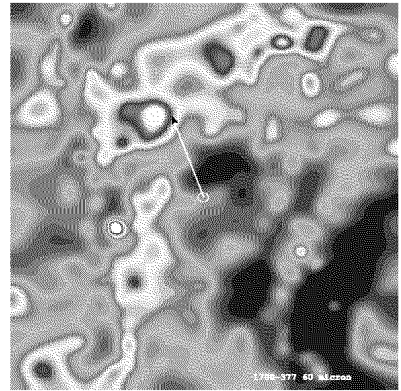,height=7.5cm}}
\caption[]{{\tt HIRAS} image of the area on the sky centered at
1700-377 at 60~$\mu$m. Many infrared sources are located in the
vicinity of 1700-377, making it difficult to identify a possible wind
bow-shock feature associated with this HMXB. East is to the left and
North is up, the image size is $0.5^{\circ} \times 0.5^{\circ}$.}
\label{fig1700}
\end{figure}

\begin{figure}[t]
\centerline{\psfig{figure=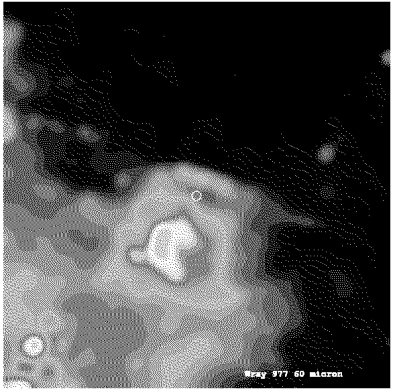,height=7.5cm}}
\caption[]{{\tt HIRAS} image of GX301-2 at 60~$\mu$m. Infrared
emission is detected which could be attributable to a wind
bow-shock. However, this is a complicated area on the sky, many other
infrared sources are located within the field of view. East is to the
left and North is up, the image has a size of $0.5^{\circ} \times
0.5^{\circ}$.}
\label{figwray}
\end{figure}

\begin{figure}[t]
\centerline{\psfig{figure=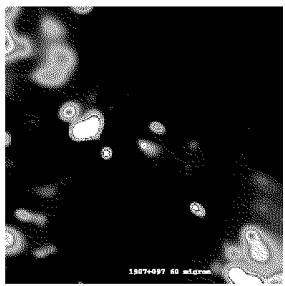,height=8cm}}
\caption[]{{\tt HIRAS} image of the HMXB 1907+097 at
60~$\mu$m. Possibly this system is producing a wind bow-shock, but this
could be a chance superposition. Infrared emission was detected in all
four IRAS bands. Note that the proper motion of 1907+097 is not
known. East is to the left and North is up, the image has a
size of $0.5^{\circ} \times 0.5^{\circ}$.}
\label{fig1907}
\end{figure}

For each system {\tt HIRAS} images were prepared in the four IRAS
bands. A wind bow-shock has been detected in a narrow-band H$\alpha$
image of the field surrounding the B-supergiant HD77581, the companion
of the X-ray pulsar Vela~X-1 (Kaper et al. \cite{kaper97}). The
bow-shock is clearly present in the {\tt HIRAS} images
too. Figure~\ref{figvela} shows a {\tt HIRAS} image of Vela~X-1 at
60~$\mu$m, with the direction of its peculiar motion indicated. We
prepared images in the other three IRAS bands (12, 25 and 100~$\mu$m,
not shown here), but the bow-shock (and at 60 and 100~$\mu$m the
H~{\sc ii} region surrounding the system) is detected in all four
bands. The arc-like structure of the bow-shock is resolved at 25 and
60~$\mu$m. Images obtained with ISOCAM onboard the {\it Infrared Space
Observatory} confirm the detection at 6 and 12~$\mu$m, and reveal the
fine structure present in the H$\alpha$ image (Kaper et
al. \cite{kaper98}). The symmetry axis of the bow-shock is parallel to
the system's direction of motion; its space velocity is 45~\kms\
(based on {\it Hipparcos} observations; note that Kaper et
al. \cite{kaper97} wrongly quote 90~\kms).  Moreover, the infrared
emission of the shockfront is coincident with the region producing
H$\alpha$ emission.

Based on the size of the H~{\sc ii} region surrounding HD77581, Kaper
et al. (\cite{kaper97}) derive that the number density of the ambient
medium is $\sim$10~cm$^{-3}$, which is about ten times higher than
predicted with the model of Dickey \& Lockman (\cite{dickey}) and a
distance of 120~pc above the galactic plane. The higher density is
also consistent with the observed stand-off distance, which is less
than predicted (Table~\ref{tablehmxb}).   

None of the fields we inspected show evidence for arc-like structures
that can be associated with a wind bow-shock around any of the other
HMXBs. It is not surprising that the three Be/X-ray binaries, which do
not drive strong stellar winds, do not produce a bow-shock. On
average, these systems also have lower space velocities (Chevalier \&
Ilovaisky \cite{chevalier}); this is consistent with the predictions
of models describing the evolution of massive binary systems (Van den
Heuvel et al. \cite{vdheuvel00}). However, a few OB-supergiant systems
do show some interesting features that are worthwhile
discussing. Figures~\ref{figcyg} - \ref{fig1907} show {\tt HIRAS}
images of some of these systems. The size (and direction) of the
depicted arrow is proportional to the peculiar space velocity of the
system (the size of the arrow corresponds to the displacement over the
next 60,000 years).

In three cases it remains unclear whether the observed HMXBs are
producing bow-shocks. In the case of 1700-377 and GX301-2 bright
objects are present in the field possibly hampering the identification
of a bow-shock (Figs.~\ref{fig1700} and \ref{figwray}, 60~$\mu$m
images). GX301-2 is surrounded by a faint ring-like structure at an
angular distance of $\sim 2'$ towards the north. As we do not know the
proper motion of the system it is hard to say whether this structure
is related to GX301-2. 1907+097 shows some emission in all four
IRAS-bands at the location of the source, a bow-shock might be
present. At a distance of $\sim 4$~kpc the shockfront is expected to
have an angular separation of $\sim 0.5'$, consistent with what is
observed.

In summary, only one of the HMXBs in our sample is, beyond any doubt,
associated with a wind bow-shock. The question remains why the other
systems do not produce a (detectable) bow-shock? They are moving at
high velocity and have strong stellar winds powering an X-ray
source. Unfortunately, our sample is too small to perform a
statistical study to determine why wind bow-shocks are absent. The
sample of OB runaways studied by Van Buren et al. (\cite{vanburen95})
is much better suited for such an investigation.

\section{Wind bow-shocks around supposedly single OB-runaway stars}

Van Buren et al. (\cite{vanburen95}) conducted an extended search for
wind bow-shocks around OB runaways. Based on pre-{\it Hipparcos}
proper-motion and radial-velocity measurements a total of 188
candidate runaway stars was compiled from the catalogs of Garmany et
al. (\cite{garmany}), Cruz-Gonzales et al. (\cite{cruz}) and the
entire list by Stone (\cite{stone}).  Van Buren et
al. (\cite{vanburen95}) correctly point out that this list should not
be considered systematic or complete since it inherits the biases of
the input catalogs. Among that sample, 58 candidate bow-shocks were
identified based on low-resolution images from the {\it IRAS Sky
Survey Atlas} (ISSA). In a follow-up study by Noriega-Crespo et
al. (\cite{noriega}) the 58 candidate bow-shocks were more closely
examined. They analyzed 60~$\mu$m IRAS high-resolution (HiRes) images
of these objects that were produced with a maximum correlation
algorithm (Fowler \& Aumann \cite{fowler}), a method similar to the
one we discussed in the previous section. Objects were classified as being
either (i) \emph{bow-shock producing objects}, or, (ii) if a
shockfront detection was inconlusive but the object showed excess
infrared radiation, as \emph{excess IR objects}. Three objects did
not fall in any of these catagories and were rejected (or rather,
classified as non-detections).

As mentioned above, Van Buren et al. (\cite{vanburen95}) concluded that
about a third of the 188 OB runaways are associated with a bow-shock,
i.e. the majority of the systems do not produce an (observable) wind
bow-shock. Our aim is to determine whether the properties of OB
runaways {\it with} a bow-shock differ from those for which a bow-shock
seems to be absent. We investigate the spatial distribution and the
kinematical properties of the OB runaways, their galactic height
distribution, and their spectral types.

\subsection{Properties of Van Buren's OB-runaway sample}

We checked the \emph{Hipparcos} database and found that 168 stars from
Van Buren's total of 188 candidate runaways are included. We omitted
stars with spectral type other than O or B and rejected all objects
located at distances beyond 4~kpc because of the large uncertainty in
proper motion. As a result we were left with a sample of 148 candidate
runaways, to which will be refered to as ``Van Buren's
sample''. Radial velocities were taken from the {\tt SIMBAD} database;
for objects for which no radial velocity is measured, the tangential
velocity is multiplied by a factor $\sqrt{3/2}$ when calculating the
space velocity.  Proper motions and radial velocities were corrected
for differential galactic rotation and the peculiar solar motion. We
followed the approach described in Moffat et al. (\cite{moffat}) which
uses a flat galactic rotation curve.

\subsubsection{Spatial distribution}

\begin{figure}[t]
\centerline{\psfig{figure=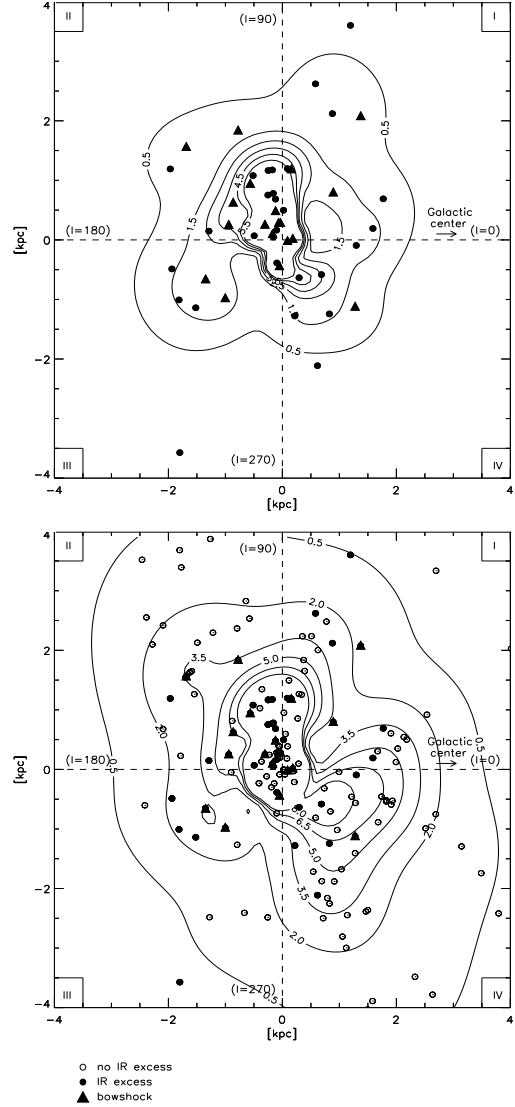,height=15cm}}
\caption[]{The spatial distribution of the OB runaways in Van Buren's
sample (bottom panel). The Sun is located at the origin of the
diagram. The density contours were calculated with the so-called
kernel method: each location of a runaway is replaced by a normal
distribution with a standard deviation corresponding to 30\% of the
heliocentric distance. The spatial distribution of the OB-runaway
stars that produce either bow-shocks (filled triangles) or show excess
infrared emission (filled circles) is displayed in the top
panel. The space density of runaways with a bow-shock decreases
significantly with distance, most likely due to the reduced
probability of detecting a bow-shock.}
\label{figcontour}
\end{figure}

The spatial distribution of the OB runaways included in Van~Buren's
sample (and with a distance less than about 4~kpc) is shown in
Fig.~\ref{figcontour}. The top panel displays the OB runaways with a
detected wind bow-shock (filled triangles) as well as those exhibiting
excess infrared emission (filled circles). The density contours were
calculated using the \emph{binormal kernel method}, where the position
of each runaway star is replaced by a binormal distribution (i.e.\
with standard deviation $\sigma_{\rm r}$ in both $x$ and
$y$). Subsequently, the convolution of all binormal distributions
results in a continuous OB-runaway density distribution in the
galactic plane. For instance, let $d_{i}$ be the distance between an
arbitrary position $(x_{i},y_{i})$ in the galactic plane and the
location $(x,y)$ of an OB runaway:
\begin{equation}
d_{i} = \sqrt {(x - x_{i})^{2} + (y - y_{i})^{2}} \, ;
\end{equation}
then the kernel is:
\begin{equation}
K_{\rm r}(d_{i} / \sigma_{\rm r}) = \frac{1}{2 \pi} e^{-
\frac{d^{2}_{i}}{2 \sigma_{\rm r}^{2}}}
\end{equation}
where $\sigma_{\rm r}$ is an adaptive spread in the kernel
function. We chose the value of $\sigma_{\rm r}$ (the width of the
distribution) to depend on the heliocentric distance (r) to the
respective runaway:
\begin{equation}
\sigma_{\rm r} = a \times {\rm r}
\end{equation}
We varied $a$ between 0.1 and 0.5, resulting in a more continuous
distribution with increasing $a$; we adopted $a = 0.3$ to produce
Fig.~\ref{figcontour}. Even though the positional accuracy of each
runaway is considerably smaller along the tangential direction,
$\sigma_{\rm r}$ is taken to be symmetric in both directions for
reasons of simplicity. Adding all individual kernels, the density
distribution becomes:
\begin{equation}
\rho (x,y) = \sum_{i=1}^{n} \frac{1}{\sigma_{\rm r}^2} K_{\rm r}(d_{i}
/ \sigma_{\rm r})
\end{equation}
Note that each term carries a normalization factor with the dimensions
${\rm kpc}^{-2}$. The distribution of $\rho (x,y)$ thus reflects the
surface density of OB runaways.

The same method was used to derive the density distribution of
bow-shock producing runaways only (top panel Fig.~\ref{figcontour}).
It was assumed that objects that show excess infrared emission are
producing wind bow-shocks.  Fig.~\ref{figcontour} indicates that the
number of OB runaways that produce observable bow-shocks drops
significantly beyond a distance of 2~kpc. The relative abundance of
objects showing either a bow-shock or IR-excess remains nearly
constant ($\sim$ 40 \%) up to a distance of about 2~kpc. Within 0.5,
1.0, and 2.0~kpc we respectively find 13 (out of 33), 18 (44), and 39
(98) runaways with a bow-shock; this corresponds to 39, 41, and 40~\%,
respectively. Between 2.0 and 4.0~kpc we encounter a bow-shock for 9
out of 50 runaways (18~\%). Therefore, even if the sample becomes
less complete at larger distances, the roughly constant fraction of
bow-shocks among the runaways up to a distance of 2~kpc suggests that
the sampling of the runaway population producing bow-shocks becomes
incomplete beyond that distance. This might be due to the reduction in
the detection sensitivity of IRAS.

A short intermezzo: The sensitivity of IRAS is, at best, $\sim 0.2$~Jy
at 12, 25, and 60~$\mu$m, and $\sim 1$~Jy at 100~$\mu$m, per $0'.25
\times 0'.25$ or $0'.5 \times 0'.5$ pixel at 12 and 25 or 60 and
100~$\mu$m, respectively (Moshir et al. \cite{moshir89}). The highest
IRAS flux per pixel measured in the bow-shock associated with Vela~X-1
(Fig.~\ref{figvela}) is 0.03 (0.01), 0.41 (0.01), 2.59 (0.06), and
1.04 (0.21)~Jy/px at 12, 25, 60, and 100~$\mu$m, respectively; the
value between brackets is the level of the infrared background in that
direction. The heliocentric distance to Vela~X-1 is 1.8~kpc; according
to the quoted IRAS sensitivity the bow-shock would still be marginally
detectable (only at 60~$\mu$m) if the distance to Vela~X-1 is
increased by about a factor 3. In terms of integrated flux (2.8, 38,
95, and 34~Jy) the bow-shock of Vela~X-1 is relatively bright. For
comparison, the integrated flux of the ``text-book-example'' bow-shock
around $\alpha$~Cam (distance 1.1~kpc) is 0.8, 45, 310, and 470~Jy
(Van Buren et al. \cite{vanburen95})\footnote{The latter values
indicate that the dust in the bow-shock around Vela~X-1 is hotter than
that around $\alpha$~Cam. The X-ray source could provide additional
heating.}. The angular extent of the bow-shock around Vela~X-1 is
$\sim 8'$; at a 3 times larger distance it would just be resolvable in
a {\tt HIRES} image. Given the strong variations in the galactic
infrared background emission, it is expected that several bow-shocks
have remained undetected in the IRAS survey. However, bow-shocks as
bright as those around Vela~X-1 and $\alpha$~Cam should be detectable
up to several kpc.

A striking feature of Fig.~\ref{figcontour} is the anisotropy of
bow-shock producing objects. A high concentration of bow-shocks and
IR-excess objects is found within 2~kpc from the Sun in quadrant II
($90^{\circ}\leq l \leq 180^{\circ}$), while in the opposite quadrant
IV ($270^{\circ}\leq l \leq 360^{\circ}$) the lowest fraction of
bow-shocks is detected. The dumbell shape apparent in the contours of
the bottom panel of Fig.~\ref{figcontour} reflects the spiral-arm
structure of the Milky Way.

\subsubsection{Space velocities}

\begin{figure}[t]
\centerline{\psfig{figure=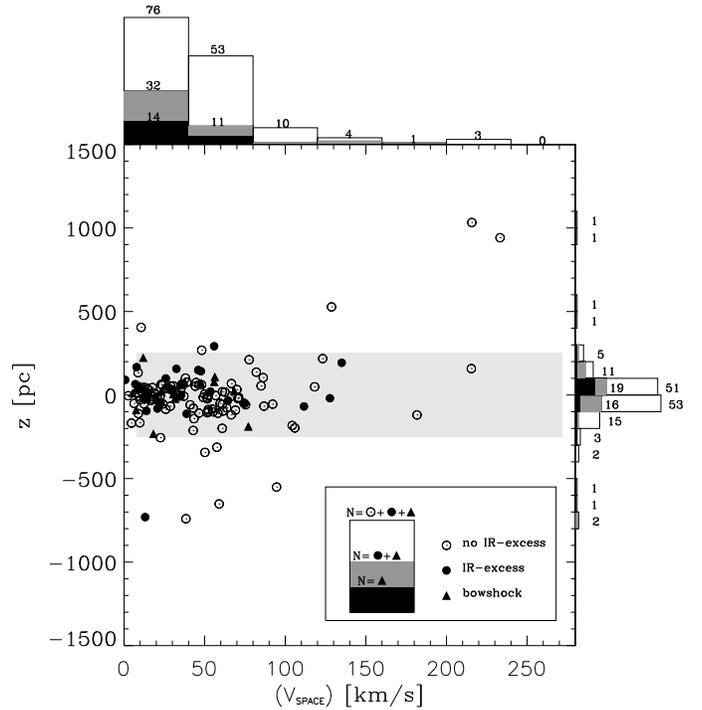,height=10cm}}
\caption[]{Space velocity versus distance $z$ to the galactic
plane. This graph shows that only a small number of objects with a
very high space velocity or with a large $z$ distance produce
bow-shocks (the meaning of the symbols is indicated in the legenda). On
the other hand, a large fraction of the objects with a space velocity
less than 40~\kms\ do show a bow-shock: 42~\% are associated
with a bow-shock or show IR excess; at higher velocities this rate is
only 22\%. The number of objects within a given space-velocity or
$z$-distance bin is indicated along the axes. The cumulative number of
objects with a bow-shock (black), with IR excess (grey), or none of the
two (white) is included.}
\label{figv}
\end{figure}

In Fig.~\ref{figv} the space velocity and distance $z$ with respect
to the galactic plane are plotted of the OB runaways in our
sample. This figure shows that most bow-shocks are detected around
objects with velocities below 40~\kms. Of the 76 objects that have a
space velocity less than 40~\kms, 14 are associated with a bow-shock
and 18 show an IR-excess, together making up for about 42\% of the
total, while among the 72 objects that are faster than 40~\kms\ a
lower fraction is found (6 bow-shocks and 10 objects with IR-excess,
resulting in a 22\% detection rate). At space velocities above
80~\kms\ none of the objects could be associated with a wind bow-shock,
and only three (out of 18) were observed with excess infrared
emission. Apparently, a very high space velocity is not a favourable
condition for the production of a wind bow-shock. This observation is
in agreement with the prediction on the basis of numerical simulations
of bow-shocks by Comer\'{o}n \& Kaper (\cite{comeron}) that at very
high velocity a shockfront becomes unstable and disrupts.

Runaways with large separations from the galactic plane also show a
low bow-shock detection rate. Among the objects that have a $z$
distance larger than 250~pc only 2 out of 13 were observed with an
IR-excess (15\% detection rate). This low detection rate of bow-shocks
at large galactic heights can be understood in terms of dust
content. Since the dust content of the interstellar medium decreases
with galactic height (Van Steenberg \& Shull \cite{steenberg}), a
runaway too far from the galactic plane may not show an observable
bow-shock because not enough dust is being swept up.

Both the space velocity and the height above the galactic plane are
physical parameters that have to be taken into account when
considering whether a bow-shock is observable or not. However, the
majority of runaways without bow-shocks are not located at extreme
distances from the galactic plane, nor have very high space velocities
(Fig.~\ref{figv}). Obviously, other factors must play an important
role as well.

\subsubsection{Dependence on spectral type}

\begin{figure}[t]
\centerline{\psfig{figure=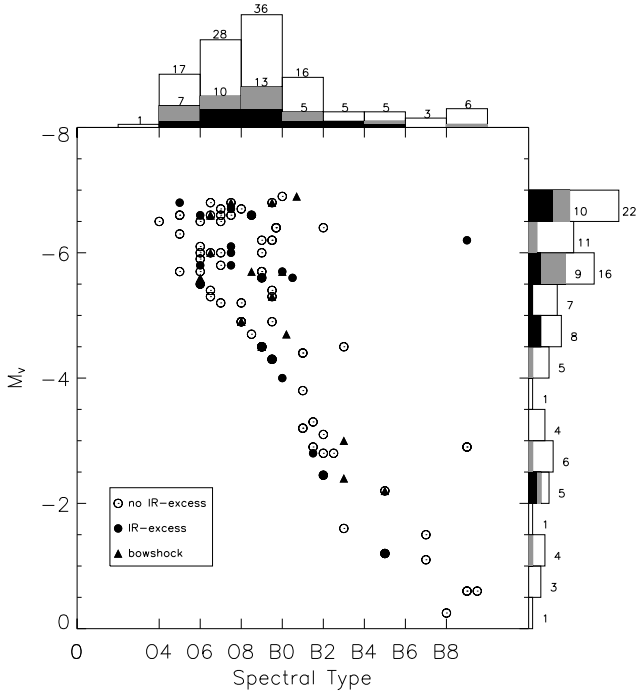,height=10cm}}
\caption[]{Hertzsprung-Russell diagram of Van Buren's sample of
runaways. Only objects located within 250~pc from the galactic plane
and with a space velocity in the range of 5 to 80~\kms\ are
included. There is no obvious correlation between the detection of a
bow-shock and the spectral type of the OB-runaway star.}
\label{fighrd}
\end{figure}

Besides the space velocity of the OB-runaway star and the density of
the ambient medium, the stellar-wind parameters \.{M} and \vinf\
determine the physical conditions under which a bow-shock is formed
(cf.\ Eq.~1). The wind mass-loss rate \.{M} is strongly dependent on
the luminosity (\.{M}$\propto L^{1.7}$, Lamers \& Leitherer
\cite{lamers93}, Puls et al. \cite{puls}), i.e. supergiants have much
stronger winds than main sequence stars.  Furthermore, the visibility
of the wind bow-shock (at infrared wavelengths) depends on the
irradiation of the swept-up dust by the ultraviolet flux from the
massive star. One would expect that supergiants are more capable of
forming a detectable bow-shock in comparison to main sequence stars.

In Fig.~\ref{fighrd} the Hertzsprung-Russell diagram of the
OB-runaway sample is presented; the symbols indicate whether a
bow-shock (or IR excess) has been detected. The OB-runaway sample
contains a relatively large fraction of supergiants and giants. This
might be due to a selection effect: {\it Hipparcos} observed only the
brightest stars on the sky. Inspection of the diagram learns that the
detection rate of bow-shocks is about 35~\%, independent of the
luminosity class of the OB runaways:
\begin{itemize}
\item From a total of 36 supergiants (luminosity class I, Ia, Ib, Iab
and II), six objects show bow-shocks and seven show excess infrared
emission (36~\% detection rate).
\item The sample contains a total of 44 giants (luminosity
class III and IV). Among these, seven show bow-shocks, and seven
objects show excess infrared emission (32~\% detection rate).
\item Out of 38 main sequence stars (luminosity class V), there are six
with bow-shocks and eight with infrared excess (37~\% detection
rate).
\end{itemize}

Also the effective temperature of the star does not seem to affect the
detectability of a wind bow-shock.

\section{On the absence of wind bow-shocks around runaway objects}

Although we found some observational indications that the presence of
a wind bow-shock depends on the space velocity of the OB runaway and
its height above the galactic plane, these findings cannot explain why
the majority of systems do not produce a wind bow-shock. In the
following we explore the possibility that perhaps some runaways are
moving subsonically, despite their high space velocity. With space
velocities found of several tens of \kms\ this would require extreme
conditions of the interstellar medium, conditions that are known to be
present in hot bubbles and galactic coronal gas.

The (isothermal) speed of sound in the interstellar medium (ISM) has a
value of $\sim$1~\kms\ at a temperature of 100~K, $\sim$10~\kms\ at
$10^4$~K and $\sim$100~\kms\ at $10^6$~K. Runaway stars thus move
supersonically unless they travel through regions with temperatures of
the order of $10^6$~K. Such hot regions are present in the Milky Way,
the so-called hot bubbles, and appear e.g. as holes in the distribution of
neutral interstellar matter (Heiles
\cite{heiles79},\cite{heiles84},\cite{heiles98}; Shull \& Saken
\cite{shull}). The hot regions are likely produced by the combined
action of stellar winds and supernovae in a central star cluster
(i.e. an OB association), which sweeps the surrounding gas and dust
away and confines this material into dense filaments (walls). The
ambient density of the medium is thereby reduced to values $\sim
10^{-3}$~cm$^{-3}$. The temperature may reach $10^6$~K, resulting in
the production of diffuse X-ray emission, such as observed from the
Cygnus superbubble (Cash et al. \cite{cash}). The ionized gas inside
the hot bubble can also be observed at radio wavelengths, with a
synchrotron component originating near the walls. The Sun is located
in the {\it local hot bubble}, a hot medium with a deficiency of
neutral matter. The walls of this bubble are fairly well defined by
the distribution of interstellar dust (Lucke \cite{lucke}) and the
detection of interstellar absorption lines (Frisch \& York
\cite{frisch83}). Recently, Heiles (\cite{heiles98}) reported the
discovery of a nearby superbubble which may have merged with the local
bubble, forming a highly oval cavity elongated towards l$\sim$230.

\begin{figure}[t]
\centerline{\psfig{figure=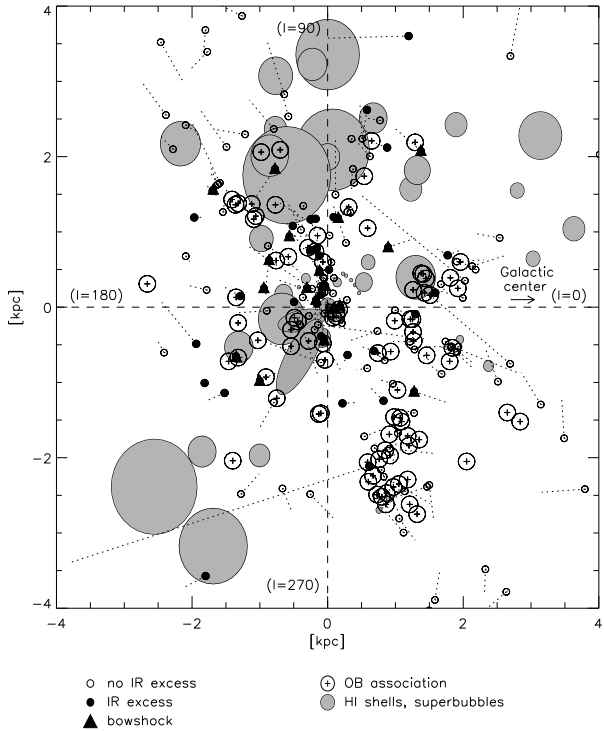,height=10cm}}
\caption[]{The distribution of H~{\sc i} shells in the galactic plane,
in the range 10$\leq$l$\leq$250 and -10$\leq$b$\leq$10 according to
Heiles (\cite{heiles79}, \cite{heiles87}) (shaded regions). The fourth
quadrant of the galactic plane has not been included in these
studies. The Sun is located in the origin of the diagram; the
direction of the galactic center is indicated. OB associations (taken
from Melnik \& Efremov \cite{melnik}) are represented by circles
including a plus sign (see legenda). OB runaways and their paths (the
path length corresponds to the projected distance travelled during the
last 10~Myr). The large concentration of OB associations in the fourth
quadrant suggests that also here hot bubbles must be present. Most of
the OB runaways producing a bow-shock do not seem to be contained
inside a hot bubble.}
\label{figh1}
\end{figure}

\begin{figure}[t]
\centerline{\psfig{figure=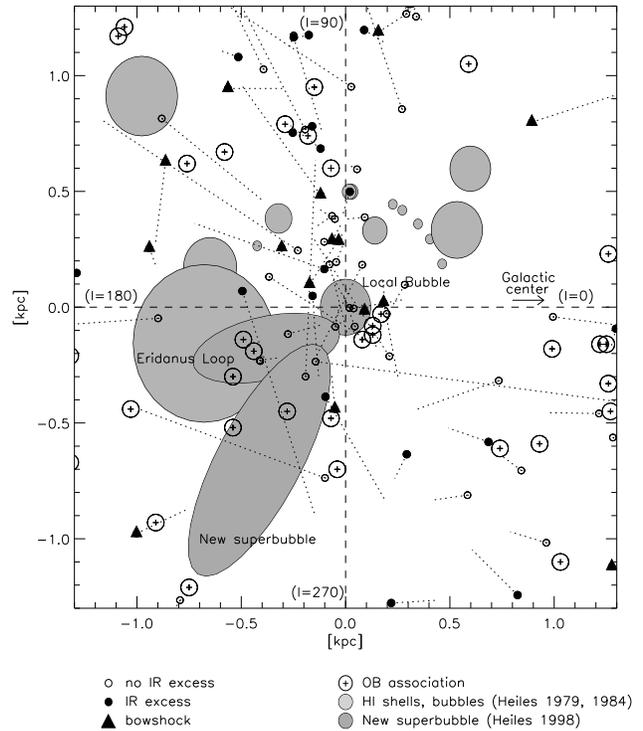,height=10cm}}
\caption[]{Zoom-in on the central area of Fig.~\ref{figh1}. Some
well-known hot bubbles are labelled.}
\label{figzoom}
\end{figure}

To test the hypothesis that such hot interstellar gas indeed prevents
runaways from producing wind bow-shocks, we investigated the
distribution of runaways in relation to the distribution of hot
bubbles in the galactic plane. We used the H~{\sc i} surveys
carried out by Heiles (\cite{heiles79}, \cite{heiles84}), and compiled
a list of H~{\sc i} shells, supershells and other shell-like
objects. The surveys span a range in galactic longitude $10^{\circ}
\leq l \leq 250^{\circ}$ and latitude $-10^{\circ} \leq b \leq
10^{\circ}$. The recently discovered nearby superbubble (Heiles
\cite{heiles98}), next to the Local Bubble, is also included. The
studies by Heiles do not cover the fourth quadrant of the galaxy
($270^{\circ} \leq l \leq 360^{\circ}$). The large number of OB
associations in that area make it very likely that hot bubbles are
present there as well. At longitudes 317$^{\circ}$ and 327$^{\circ}$
filaments extending over several degrees are detected in the
Australian Mononglo (MOST) survey (Green et al. \cite{green}).

Figures~\ref{figh1} and \ref{figzoom} show the locations and
trajectories of runaways (the length of the track corresponds to the distance
travelled in 10~Myr), projected onto the galactic plane,
with respect to H~{\sc i} shells and bubbles. Figure~\ref{figzoom} (a
zoom in on the central region in Figure~\ref{figh1}), focusses on the
solar neighbourhood (within 1.3~kpc). OB associations are represented
as circles with a cross in the middle (see legenda); in
Fig.~\ref{figzoom} the associations comprised in De~Zeeuw et
al. (\cite{zeeuw}) are included. Apart from Heiles' New Superbubble
and the Eridanus Loop, the shape of the H~{\sc i} shells is assumed to
be spherically symmetric. The estimated distance and angular size in
galactic latitude and longitude are used to determine the radius of
the bubble. A large degree of uncertainty exists in the size of the
bubbles along the line of sight.

Although both the OB-runaway and the hot-bubble distribution are not
accurately known, the OB runaways producing bow-shocks are mostly found
in regions devoid of hot bubbles. Remarkable is the asymmetry in the
distribution of OB runaways producing bow-shocks with respect to the Sun:
quadrant II (in the direction of Cepheus) contains significantly more
bow-shocks than the other quadrants. This anisotropy agrees with the
distribution of hot bubbles. In the nearby (super)bubbles we hardly
encounter any OB runaway with a bow-shock.

Dove \& Shull (\cite{dove}) estimate that the average volume filling
factor of (super)bubbles and cavities in the galactic plane is
50--80\%, which is consistent with the fraction of bow-shock producing
objects among the runaway stars in Van~Buren's sample. Apparently, the
soundspeed in these regions is high enough to prevent the formation of
wind bow-shocks.

\subsection{The HMXB sample}

\begin{figure}[t]
\centerline{\psfig{figure=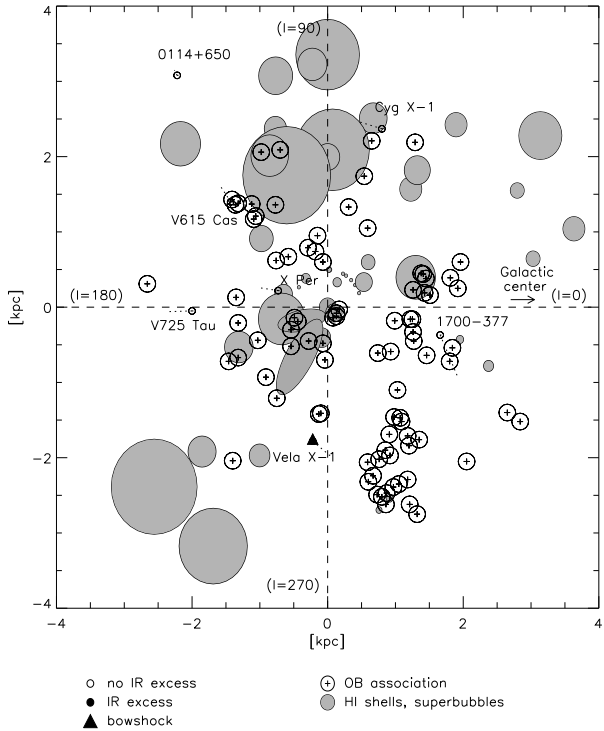,height=10cm}}
\caption[]{The distribution of the HMXBs in our sample is shown with
respect to hot bubbles and OB associations. Each HMXB shows a track
that reflects the distance travelled in 10~Myr; in some cases these
tracks identify the parent OB association of the system. Cyg~X-1 and
X~Per are very likely enclosed in hot superbubbles. Also 1700-377 and
V615 Cas may be situated in such hot regions, although the presence of
bubbles at the positions of 1700-377 and V615 Cas is uncertain.}
\label{fighmxb}
\end{figure}

Can the detection of a wind bow-shock around only one HMXB (out of 11)
also be explained by the distribution of hot bubbles?
Figure~\ref{fighmxb} displays the distribution of the HMXBs in our
sample.  Cyg~X-1 is located close to the edge of the Cygnus
superbubble, while Vela~X-1, the only HMXB for which a bow-shock
clearly has been detected, seems to move through a medium devoid of
hot bubbles. At least four HMXBs are likely to be travelling through a
hot bubble.

{\bf 1700-377}: Benaglia \& Cappa (\cite{benaglia}) report
the presence of an H~{\sc i} structure that may be due to a wind-blown
bubble created by the O-supergiant companion of 1700-377, HD 153919. The
H~{\sc i} region increases in size towards higher galactic latitudes,
forming some kind of horseshoe shape. The mean velocity of the bubble
corresponds to a distance $\sim$2~kpc, in agreement with the
distance of 1700-377 (cf.\ Ankay et al. \cite{ankay}).

{\bf V615 Cas.} According to Figure~\ref{fighmxb}, this HMXB is
located near the edge of a large superbubble; two other bubbles are
found at roughly the same longitude. Shell GS 135+08-34 has a larger
catalogued distance than V615 Cas (3.1 and 2.0~kpc, respectively), but
taking into account the uncertainty in these distance determinations,
they could possibly be associated with each other. On the other hand,
the center of GS 135+08-34 lies at a larger angular separation from
the galactic plane than V615 Cas ($+8^{\circ}$ as compared to
$+1^{\circ}$). It is not quite clear whether the extent of GS
135+08-34 is sufficient to enclose V615 Cas. Heiles (\cite{heiles84})
estimates an angular size of only $\sim 6^{\circ}$ of GS 135+08-34 in
galactic latitude.

{\bf X Per.} Two H~{\sc i} shells can be discerned in the vicinity of
X~Per. The smaller of the two bubbles (GS 165-21-4) is centered around
galactic latitude b$=-21^{\circ}$ corresponding to a separation from
the galactic plane of $\sim 260$~pc at a distance of 0.72~kpc (Heiles
\cite{heiles84}). The extent of this shell along the $z$-direction is
estimated to be $\sim$160~pc at that distance. X~Per is situated 235~pc
below the galactic plane, and is therefore likely enclosed in this
bubble.

{\bf Cyg X-1.} This HMXB is located in the boundary region of the
Cygnus superbubble (see Figure~\ref{figcyg}). Cash et al. (1980)
performed observations with HEAO-1 of this region and showed that the
observed X-rays emitted from this region is outlined by a ring of
elongated H$\alpha$ filaments, suggesting a common origin. Moreover,
interstellar absorption considerations yield a distance of about 2~kpc
of the X-ray feature, thus at roughly the same distance as Cyg
X-1. The temperature of the X-ray emitting plasma contained in the
Cygnus superbubble is estimated to be 2$\times 10^6$~K.


\begin{figure}[t]
\centerline{\psfig{figure=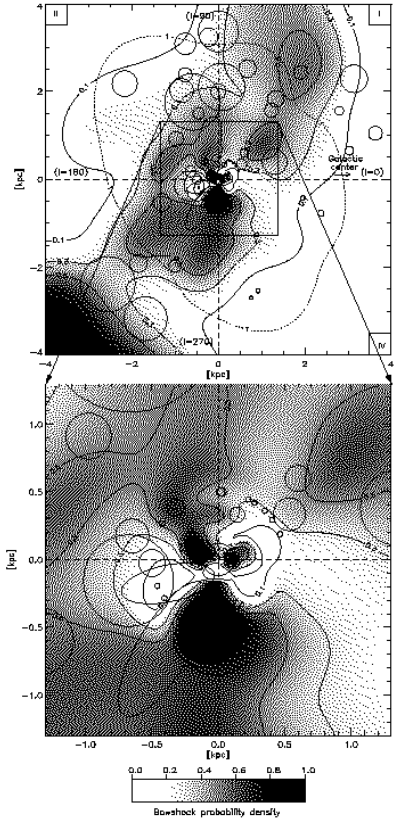,width=9cm}}
\caption[]{The fraction of bow-shock producing runaways (``bow-shock
probability density'') is shown together with the observed
distribution of local hot bubbles. The bow-shock probability
distribution is determined with objects that are located within 300 pc
from the galactic plane and have space velocities between 5 and
80~\kms. The HMXB sample is also included, yielding a total sample of
137 runaways. In the top panel, the dotted contour follows a runaway
number density of 1~kpc$^{-2}$, marking the region outside of which
the bow-shock probability density is determined by less than 1 object
per square kpc.}
\label{figdensity}
\end{figure}

\begin{figure}
\centerline{\psfig{figure=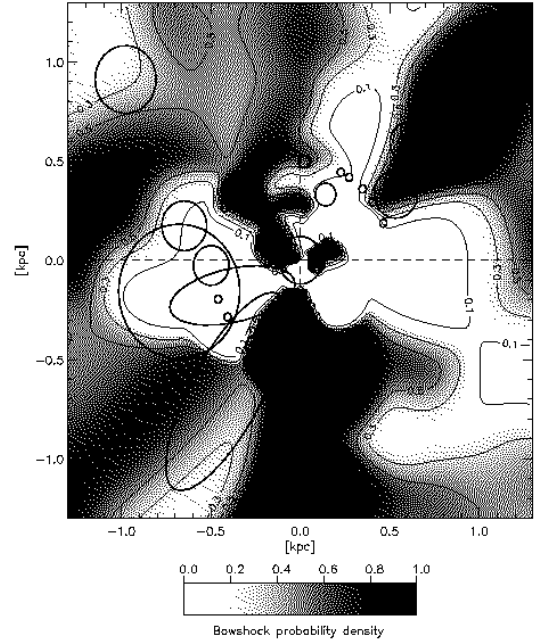,width=9cm}}
\caption[]{As the bottom panel of Fig.~\ref{figdensity}, but now
calculated adopting $\sigma_{\rm r} = 0.15 r$ to illustrate that the
obtained space distribution is not very sensitive to the value of
$a$.}
\label{figdensity2}
\end{figure}

\section{Discussion}

One of the main conclusions of our work is that the formation of a
wind bow-shock around an OB-runaway star strongly depends on the
temperature and density of the ambient medium. Apparently, the
majority of the OB runaways are moving through hot bubbles distributed
in the galactic plane. These hot bubbles are created by the combined
action of stellar winds and supernovae of OB stars, most of them
members of OB associations. Assuming that all OB runaways originate in
OB associations, it is very likely that they travel through such hot
bubbles for a substantial fraction of their (relatively short)
lifetime. In the following we use the absence of a wind bow-shock as a
diagnostic to probe the distribution of hot bubbles in the solar
neighbourhood. Finally, we present arguments why HMXBs should have a
higher probability to be located inside a hot bubble than dynamically
ejected OB runaways.

\subsection{The distribution of hot bubbles in the solar
neighbourhood} 

Various people have tried to map out the extent of the local bubble
(see Frisch \cite{frisch95} and references therein). Due to its high
temperature, some nearby runaways that are moving through this bubble
are not producing wind bow-shocks, even though they have substantial
space velocities. The presence, or rather the absence, of bow-shocks
can thus serve as a diagnostic to constrain the extent of the local
bubble. An attempt to map out hot galactic regions based on the
relative presence of bow-shock producing runaways is shown in
Fig.~\ref{figdensity}. The {\it bow-shock probability density}
reflects the fraction of runaways producing a bow-shock. The galactic
distribution is derived using the binormal kernel method (adopting
$\sigma_{\rm r} = 0.30 r$, cf.\ Section 3.1.1); the spatial
distribution of OB runaways producing bow-shocks (top panel
Fig.~\ref{figcontour}) is divided by the spatial distribution of the
whole sample of OB runaways, including the HMXBs (bottom panel
Fig.~\ref{figcontour}). Some OB runaways can be expected to not
produce a bow-shock, because of their high space velocity or their
large separation from the galactic plane. For that reason, the objects
used for determining the probability distribution in
Fig.~\ref{figdensity} are only those with intermediate velocities
(5~\kms\ $\leq V_{\rm space} \leq$ 80~\kms), and those that are
located within 300~pc of the galactic plane. Note that areas with a
small number of runaways will have a rather inaccurate value of the
bow-shock probability density (division through a small
number). Fig.~\ref{figdensity2} illustrates that the obtained
distribution is not very sensitive to the value of $a$;
for $\sigma_{\rm r} = 0.5 r$ the space distribution becomes uniform at
distances beyond $\sim 0.7$~kpc from the Sun.

It can be seen that the distribution of nearby hot bubbles (from
H~{\i} measurements) is consistent with the distribution of regions
showing a low fraction of bow-shock producing runaways. At larger
distances this method of mapping hot regions becomes inaccurate due to
the rapidly decreasing number of objects. For example, the probability
density maximum in the lower left corner of Fig.~\ref{figdensity} (top
panel) is completely determined by only one object beyond 3 kpc in
that quadrant, an object that happens to show excess infrared
emission.

\subsection{HMXBs have a lower probability to produce bow-shocks than
dynamically ejected OB runaways}

Only one of the 11 HMXBs has an associated wind bow-shock. Though
small-number statistics, does this indicate that HMXBs have a lower
probability to form a wind bow-shock?  The bow-shock detection rate for
Van~Buren's sample is about 40~\%. If we take into account that the
detection sensitivity of IRAS is poor for objects beyond 2~kpc, then
we are left with 1 out of 5 HMXBs.  These numbers do not provide
statistically significant results, but the different evolutionary
origin, and therefore on average different kinematical age, of HMXBs
and dynamically ejected runaways might result in a difference in
detection rate of wind bow-shocks.

Massive stars ejected by dynamical interactions should have escaped
the OB association at a relatively early stage, when the cluster is
still dense and the probability for close encounters high. On the
other hand, OB stars in HMXBs went through several stages of binary
evolution before they became runaways after the occurence of a
supernova within the system. Therefore, the kinematical age (i.e. the
travel time since ejection took place) of HMXBs should be
smaller than those of dynamically ejected runaways. As a consequence,
HMXBs have a higher probability to be still enclosed in the hot and
rarefied regions (e.g. superbubbles) that surround OB associations.
Further study is required to prove this hypothesis with observations.

\section{Summary and conclusions}

Based on our study of OB runaways and the phenomenon of wind-bow-shock
formation we conclude that about 40~\% of the OB-runaway stars produce
wind bow-shocks. This fraction is a little bit higher than proposed by
Van~Buren and coworkers, but takes into account the estimated
reduction in IRAS sensitivity required to detect a bow-shock. We
investigate why the majority of OB runaways do not produce a
bow-shock. The spectral type (temperature, luminosity, mass-loss rate),
space velocity and separation from the galactic plane seem to be of
minor importance. We propose that the temperature and density of the
ambient medium are the dominant factors. In hot bubbles the speed of
sound is larger than the typical velocity of a runaway star
($\sim$50~\kms), so that runaways in these regions do not move
supersonically and would not produce a bow-shock.  OB-runaways produced
through the binary supernova scenario should on average have a shorter
kinematical age than those produced through the cluster ejection
scenario, and thus have a higher probability to be still inside a hot
bubble. For some runaways, e.g. Cyg~X-1, there is observational
evidence that the system indeed is contained in a (super)bubble.

\begin{acknowledgements}
We are grateful to Romke Bontekoe and Do Kester for the permission to
work with their software, and their help and advice. Ben Stappers, Ed
van den Heuvel, and the referee are thanked for a careful reading of
the manuscript and useful suggestions.  The IRAS data were obtained
using the IRAS data base server of the Space Research Organisation of
the Netherlands (SRON) and the Dutch Expertise Centre for Astronomical
Data Processing funded by the Netherlands Organisation for Scientific
Research (NWO).  The IRAS data base server project was also partly
funded through the Air Force Office of Scientific Research, grants
AFOSR 86-0140 and AFOSR 89-0320. We acknowledge the use of the SIMBAD
database. LK is supported by a fellowship of the Royal Netherlands
Academy of Arts and Sciences.
\end{acknowledgements}

\end{document}